\documentclass[aps,pra,superscriptaddress,reprint,floatfix,longbibliography,showpacs]{revtex4-1}
\usepackage[english]{babel}
\usepackage{amsmath}
\usepackage{amssymb}
\usepackage{graphicx}
\usepackage{hyperref}
\usepackage{upgreek}
\usepackage{epstopdf}
\usepackage{color}
\usepackage{dcolumn}
\usepackage{multirow}
\usepackage[note-name=, use-sort-key = false]{notes2bib}

\def\cm{cm$^{-1}$}

\begin{document}
\title{Evolution of the optical response of the magnetic topological insulators Mn(Bi$_{1-x}$Sb$_{x}$)$_2$Te$_4$ with Sb content}
\author{M. K\"opf}
\affiliation{Experimentalphysik II, Institute of Physics, Augsburg University, 86159 Augsburg, Germany}
\author{S. H. Lee}
\affiliation{2D Crystal Consortium, Materials Research Institute, Pennsylvania State University, University Park, PA 16802, USA}
\affiliation{Department of Physics, Pennsylvania State University, University Park, Pennsylvania 16802, USA}
\author{Z. Q. Mao}
\affiliation{2D Crystal Consortium, Materials Research Institute, Pennsylvania State University, University Park, PA 16802, USA}
\affiliation{Department of Physics, Pennsylvania State University, University Park, Pennsylvania 16802, USA}
\affiliation{Department of Materials Science and Engineering, Pennsylvania State University, University Park, Pennsylvania 16802, USA}
\author{C. A. Kuntscher}\email{christine.kuntscher@physik.uni-augsburg.de}
\affiliation{Experimentalphysik II, Institute of Physics, Augsburg University, 86159 Augsburg, Germany}

\begin{abstract}
MnBi$_2$Te$_4$ is a promising representative of intrinsic antiferromagnetic topological insulators, which could enable rare quantum mechanical effects like the quantum anomalous Hall effect. Especially at low temperatures, numerous studies have been reported, demonstrating the great potential of this compound in the magnetically ordered state below $T_{\mathrm{N}}$. Among recent findings, the alloy compound Mn(Bi$_{1-x}$Sb$_x$)$_2$Te$_4$ has been suggested to be an interesting candidate for the realization of an ideal Weyl semimetal state. By exchanging Bi by Sb, the electronic structure is influenced in terms of a shift of the Fermi energy, and a decrease in the energy gap has been predicted. In this work, we investigate and compare the optical conductivity of Mn(Bi$_{1-x}$Sb$_x$)$_2$Te$_4$ single crystals with various Sb doping levels $x$ by infrared reflectivity measurements. We observe a big impact of the Sb content on the low-energy excitations characterizing the metallic state of our samples. Different $x$=0.26 crystals show significant differences in their optical response and also a strong position dependence. The findings are compared to the recently reported optical conductivity spectrum of the pure compound MnBi$_2$Te$_4$.

\end{abstract}

\maketitle

\section{Introduction}
The material MnBi$_2$Te$_4$ (MBT) has been established as a highly interesting material comprising several topological quantum states, such as antiferromagnetic topological insulator, ferromagnetic type-II Weyl semimetal, as well as axion insulator and quantum anomalous Hall insulator phases in thin films, depending on the number of layers~\cite{Deng.2020}. In particular, first-principles calculations showed that the electronic band structure of the ferromagnetic phase of MBT includes only a single pair of tilted Weyl cones, hence hosting a minimal ideal type-II Weyl semimetal state~\cite{Swatek.2020}.
MBT belongs to the group of ternary chalcogenides and is characterized by a van der Waals-type layered structure with $R\bar{3}m$ space group, consisting of Te-Bi-Te-Mn-Te-Bi-Te septuple layers~\cite{Li.2020, Lei.2021}.
The exotic quantum effects in MBT can be ascribed to the combination of topological surface states originating from the Bi$_2$Te$_3$ components and the magnetic contribution of Mn$^{2+}$ ion. MBT is in a paramagnetic state above the N\'{e}el temperature $T_{\mathrm{N}}=24\,$K~\cite{Chen.2019a}. Below $T_{\mathrm{N}}$ it shows $A$-type antiferromagnetic ordering, where the magnetic moments of the Mn$^{2+}$ ions are aligned parallel within the $ab$-plane and anti-parallel along the $c$-direction~\cite{Li.2020}.

The gradual exchange of Bi by Sb to obtain a Mn(Bi$_{1-x}$Sb$_x$)$_2$Te$_4$ (MBST) alloy compound recently emerged as a more interesting possibility to achieve an ideal Weyl semimetal state~\cite{Lee.2021}: For samples with Sb doping level $x$$\approx$0.26 it was shown that the chemical potential is close to the conduction/valence band edge, and that even samples from the same batch could show lightly electron doped or lightly hole doped transport characteristics. According to the magnetic field dependence of the Hall resistivity, the $x$$\approx$0.26 samples have a distinctly lower carrier density and a higher carrier transport mobility in the paramagnetic phase as compared to other compositions. In high magnetic fields along the $c$ axis and at low temperature the field-induced antiferromagnetic-to-ferromagnetic transition in MBST with $x$$\approx$0.26 drives a type-II Weyl semimetal state with the coexistence of electron and hole pockets \cite{Lee.2021}. As the Sb-doping appears to be a promising route for obtaining new topological phases in MBT, further studies on the properties of Mn(Bi$_{1-x}$Sb$_x$)$_2$Te$_4$ materials as a function of Sb content are of high relevance.

By increasing the Sb doping level, the crystal structure of Mn(Bi$_{1-x}$Sb$_x$)$_2$Te$_4$ is barely changing, as the cell parameters amount to $a = 4.33\,$\AA\  and $c = 40.93\,$\AA\ for pure MBT, as compared to $a = 4.25\,$\AA\  and $c = 40.87\,$\AA\ for MnSb$_2$Te$_4$~\cite{Yan.2019a}. On the other hand, rather pronounced changes in the electronic structure of MBST have been reported for different Sb doping levels. Starting as an $n$-type doped metal for the pure compound MBT, the material undergoes an electronic transition to $p$-type at a doping level $x$$\approx$0.26~\cite{Chen.2019,Yan.2019a,Lee.2021}. Thus, the chemical potential of the $x$$\approx$0.26 compound is expected to be close to the valence or conduction band edges, as both cases have been reported for several samples by Lee {\it et al.}~\cite{Lee.2021}. With increasing Sb doping level, the metallic $p$-type character is steadily rising~\cite{Chen.2019}. In addition, the energy gap is changing its value from -188\,meV for MBT to 34\,meV for MnSb$_2$Te$_4$. Due to the sign change, a topological transition point has been predicted at $x=0.55$ via first-principle calculations by Chen {\it et al.}~\cite{Chen.2019}. Therefore, it was suggested that MBST is losing its topological properties, as the band gap is re-inverted at $x=0.55$, which results in a topologically trivial insulator for doping ratios $x>0.55$~\cite{Chen.2019,Ko.2020}. Yet, some studies also have found topological characteristics for MnSb$_2$Te$_4$ ($x=1$) with an energy gap of 16\,meV, which is in contradiction to the results by Chen {\it et al.}~\cite{Chen.2019}.
The magnetic properties change with the Sb content as well: With increasing $x$, the magnetic phase transition temperature is decreasing, i.e., from $T_{\mathrm{N}}=24\,$K for $x=0$ to $T_{\mathrm{N}}=19\,$K ($x=1$), and the coercive magnetic field, the saturation and the Weiss constant are all decreasing in their absolute values~\cite{Yan.2019a}. Furthermore, for high Sb doping level, the antisite mixing between Mn and Sb ions can lead to a ferrimagnetic ground state depending on the crystal growth conditions \cite{Liu.2021,Lai.2021,Riberolles.2021,Chen.2020,Murakami.2019}.

Infrared spectroscopy is a powerful technique to characterize the electronic excitations in a material with a high energy resolution. Recently, the effect of the magnetic ordering on the electronic band structure of MBT and the doped Mn(Bi$_{1-x}$Sb$_x$)$_2$Te$_4$ compound with $x=0.93$ could be observed in the temperature-dependent optical conductivity spectra \cite{Koepf.2020,Xu.2021,Koepf.2022}:
Close to the magnetic ordering temperature the low-energy excitations of the free charge carriers show an anomaly, namely a maximum in the
screened and unscreened plasma frequency.
These results could be explained by the hybridization of the $p$ bands of Bi and Te close to the Fermi level, forming the Dirac cone at the surface, and the $d$ bands related to the Mn atoms further away from the Fermi level, and this hybridization is strong enough to create an exchange gap at the topological surface states when the antiferromagnetic ordering sets in~\cite{He.2020}.
Although infrared spectroscopy is mainly bulk sensitive, also surface-related excitations can influence the optical conductivity spectra, similar to recent findings for the magnetic topological insulator EuIn$_2$As$_2$ \cite{Xu.2021a,Regmi.2020},

In this paper, we investigate the effect of Sb doping on the optical conductivity of the alloy compound MBST as obtained by infrared reflectivity measurements over a broad frequency range.
The results are related to recent findings on the electronic properties of MBST for various doping levels~\cite{Chen.2019,Yan.2019a,Lee.2021}.

\begin{figure*}[t]
	\includegraphics[width=0.95\linewidth]{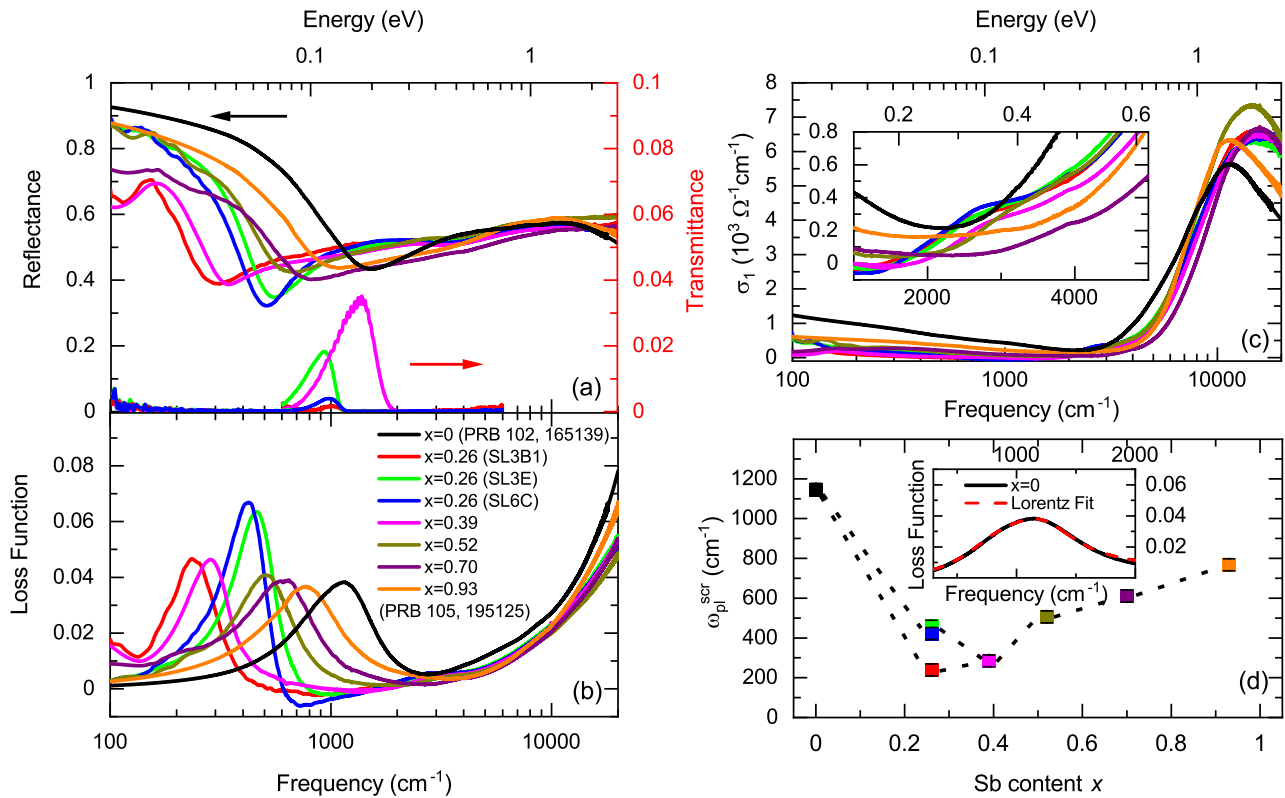}
	\caption{\label{fig.optics} (a) Reflectance and transmittance, (b) loss function and (c) optical conductivity $\sigma_1$ of eight MBST samples with various Sb doping levels $x$, namely 0 (from Ref.\ \cite{Koepf.2020}), 0.26 (labelled SL3B1, SL3E, SL6C), 0.39, 0.52, 0.70, and 0.93.
The inset of (c) highlights the frequency range around the strong onset of interband transitions.
(d) Values of the screened plasma frequency $\omega_{\mathrm{pl}}^{\mathrm{scr}}$ for each sample, as obtained from the loss function by Lorentz fitting (see text). The inset shows a Lorentz fit of the loss function for the pure compound, as an example.
Please note that the color coding of the symbols refers to the legend given in (b).}
\end{figure*}

\section{Methods}
Single crystals of MBST with doping ratios $x=0, 0.26, 0.39, 0.52, 0.70, 0.93$ were grown by the self-flux method as reported in Ref.~\cite{Yan.2019}. The MBST crystals have been characterized in detail by temperature-dependent resistivity and Hall resistivity measurements reported in Ref.\ \cite{Lee.2021}.
The plate-like samples had surface areas of 0.5\,mm $\times$ 1\,mm and the thicknesses varied from 20\,$\upmu$m up to 200\,$\upmu$m.
We performed room-temperature reflectance measurements with the use of a Bruker Hyperion infrared microscope, which is coupled to a Bruker Vertex80v FTIR spectrometer. The data have been collected from the far-infrared up to the visible range (100 to 20000\,\cm). As reference we have used an unprotected aluminum mirror, and both the sample and the reference mirror have been carefully aligned perpendicular to the incoming infrared beam. In addition, we also measured the transmittance of our samples on the same position as the reflectance measurement, where the empty beam path was used as reference. The reflectance spectra were combined and extrapolated in the low- and high-frequency range with the help of literature values and volumetric data. After this procedure, we applied the Kramers-Kronig relations in order to calculate optical functions such as the real part of the optical conductivity $\sigma_1$, the real part of the dielectric function $\varepsilon_1$ and the loss function -Im(1/$\epsilon$), using programs by David Tanner~\cite{Tanner.2015}. The optical spectra were fitted with the Drude-Lorentz model using the software RefFIT~\cite{Kuzmenko.2005}.

\begin{figure}[t]
	\includegraphics[width=0.9\linewidth]{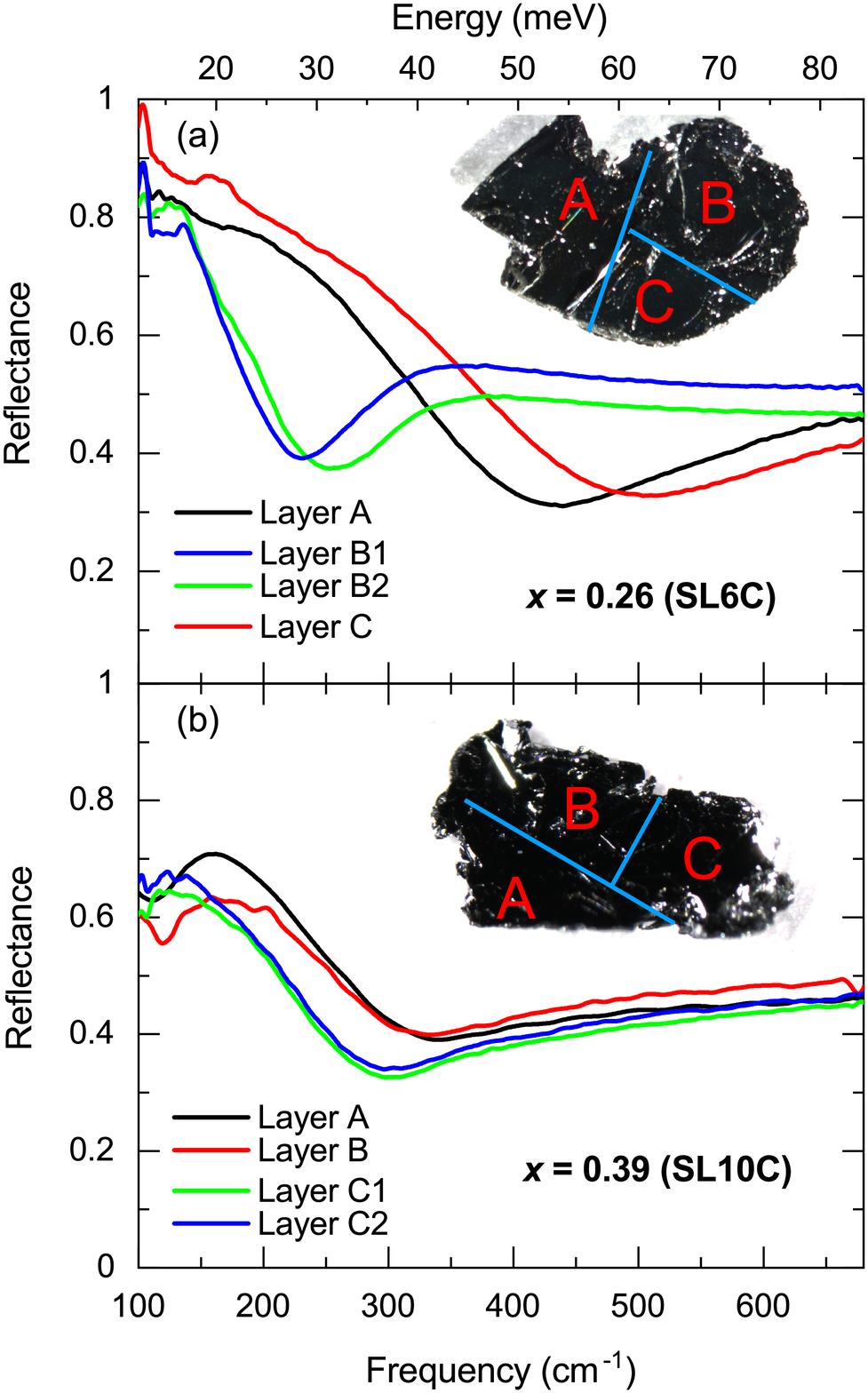}
	\caption{\label{fig.positions} Position/Layer dependence of the reflectance in the low-energy region of the (a) SL6C $x=0.26$ sample and the (b) SL10C $x=0.39$ sample. The insets of (a) and (b) display the measured crystals with the lateral dimensions 2.5\,mm $\times$ 1.5\,mm and 1\,mm $\times$ 0.8\,mm, respectively. The diameter of the probing spot size amounted to 200\,$\mu$m.}
\end{figure}

\begin{figure*}[t]
	\includegraphics[width=0.95\linewidth]{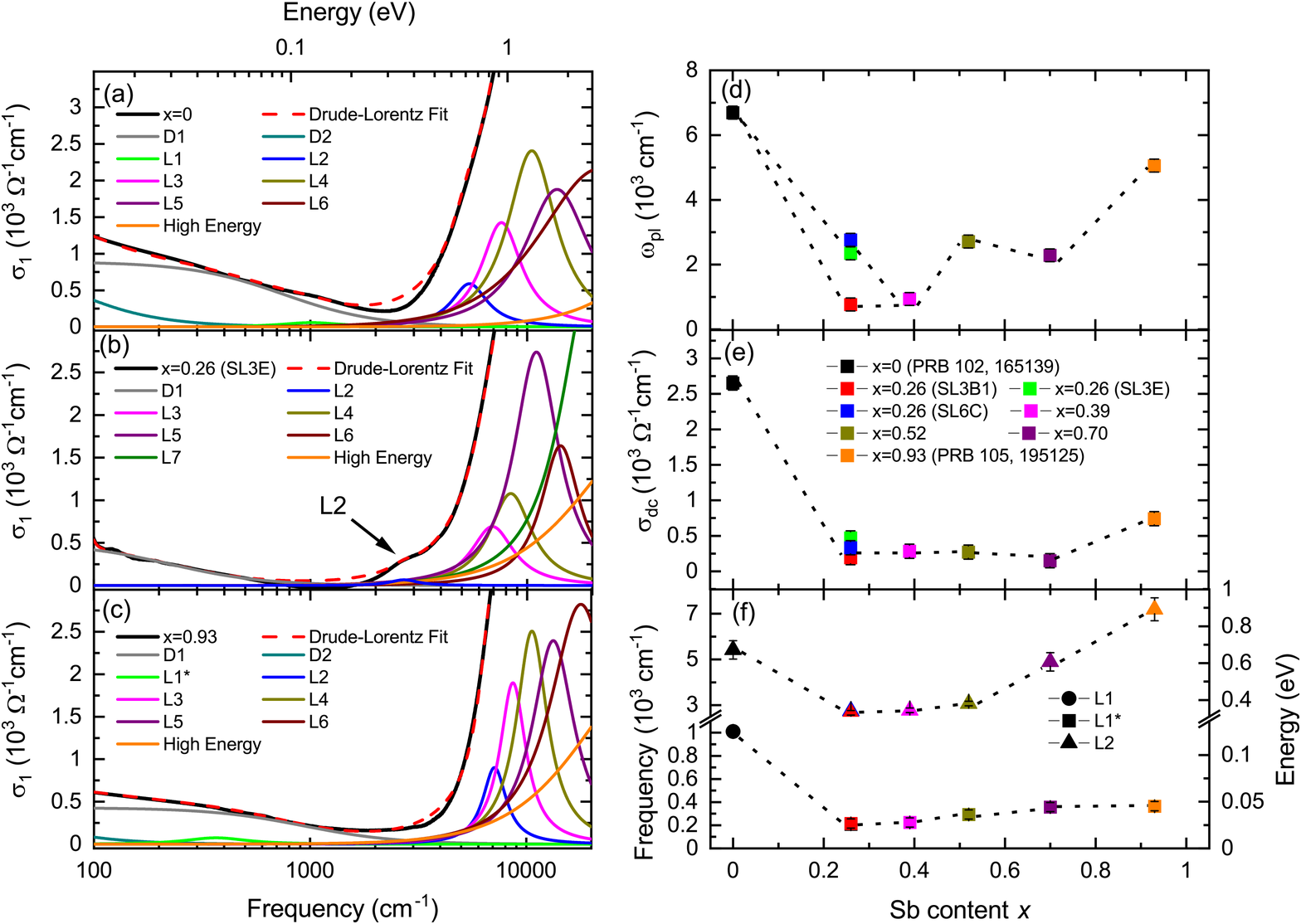}
	\caption{\label{fig.fitting} Drude-Lorentz fits of the optical conductivity, including all Drude and Lorentz contributions, for the (a) $x=0$, (b) $x=0.26$ (SL3E) and (c) $x=0.93$ sample.
	The corresponding parameters resulting from these fits, namely the plasma frequency $\omega_{\mathrm{pl}}$ (see text), the dc conductivity $\sigma_{\mathrm{dc}}$,
 and the position of the Lorentz functions L1, L1* and L2, are plotted in (d), (e) and (f), respectively.}
\end{figure*}

\section{Results and Discussion}

The reflectance spectra for MBST samples with various Sb doping levels $x$ are shown in Fig.\,\ref{fig.optics}\,(a). Clearly, these samples show different profile characteristics in the spectra below $\sim$2000\,\cm, whereas above 2000\,\cm\ the changes with $x$ are much less pronounced. Accordingly, the Sb doping has a much stronger influence on the low-energy excitations. As expected, the metallic character is strongly affected by the Sb doping level. Starting with the pure compound MBT, which has been measured and published earlier \cite{Koepf.2020,Xu.2021}, we observe the largest free charge carrier contribution with a high reflectance above 90\,\% at 100\,\cm\ and a plasma edge, both signaling the metallic character.
For 26\,\% doping, three different samples SL3B1, SL6C and SL3E were measured. These exhibit a strongly reduced metallicity as compared to MBT, with a shift of the plasma edge to low frequencies. This reduced metallic character is in agreement with magnetotransport measurements \cite{Lee.2021}. Despite the identical nominal composition of the three samples, there are significant differences in the profile of their reflectance spectra: While samples SL6C and SL3E show a plasma edge close to 500~cm$^{-1}$, sample SL3B1 has a much weaker metallic contribution with a plasma edge near 200\,\cm. Surprisingly, the reflectance spectrum of the 39\,\% doped sample looks very similar to that of the SL3B1 sample with $x$=0.26. For Sb doping levels above 39\,\%, the metallic character is growing with increasing Sb content, as revealed by the shift of the plasma edge and plasma minimum to higher energies.
Thus, considering the reflectance of the measured samples, we find a good agreement of our data and the results of Chen {\it et al.}~\cite{Chen.2019}.
It is important to note that solely based on the optical data, we cannot discriminate between electron- or hole-type free charge carriers present in the materials.
The corresponding transmittance spectra are depicted in Fig.\,\ref{fig.optics}\,(a) as well. Only the weakly metallic samples, such as the 26 and 39\,\% Sb doped samples, which are also thin at the same time, are slightly transparent and only over a small frequency range between approx.\ 600 and 2000\,\cm.
The partial transparency of the samples causes small Fabry-P\'erot interference fringes in the reflectance spectra.

In order to check, whether the surprising results for the 39\,\% doped sample could be related to inhomogeneities of Sb doping in the studied crystal, we carried out position-dependent reflectivity measurements in the low-energy range, where the largest spectroscopic differences among the samples were observed. We measured several samples, including the $x=0.39$ one, with a probing spot size 200\,$\mu$m. The results for two samples as representative examples are
plotted in Fig.\,\ref{fig.positions}.
The 26\,\% doped SL6C sample [Fig.\,\ref{fig.positions}\,(a)] shows a relatively strong layer/position dependence, as the excitations on layer B differ a lot from those of layer A and C. Therefore, it seems that even though the position dependence of the doping level is negligible (below 1\,\%), as shown by energy dispersive x-ray spectroscopy (EDX) measurements, there is a significant effect on the optical response for this composition.
In contrast, the 39\,\% doped sample, which is also weakly metallic, shows a much smaller layer dependence [see Fig.\,\ref{fig.positions}\,(b)].
Thus, our results are in good agreement with those in Ref.\ \cite{Lee.2021}, as several samples, partially from the same batch and with the same ``critical'' doping $x=0.26$, turned out to possess slightly different charge carrier densities and also different types of free charge carriers.

The loss function and the optical conductivity of the samples, as obtained from the Kramers-Kronig analysis of the reflectance data, are depicted in Figs.\ \ref{fig.optics} (b) and (c), respectively.
They become slightly negative (-0.006 and -60\,$\Omega^{-1}\mathrm{cm}^{-1}$, respectively) at around 1000\,\cm\, for doping levels around $x$=0.26, which is by and large within our error bars and means that the loss function and $\sigma_1$ have a minimum close to zero.
The metallic character of the doped samples is weak, as indicated by the small value of the optical conductivity ($\sigma_1$$<$1000\,$\Omega^{-1}\mathrm{cm}^{-1}$) below $\sim$2000\,\cm.
This region is followed by a strong increase of $\sigma_1$ above 4000 - 5000\,\cm\ and a broad maximum in the frequency range 10000 - 20000\,\cm, which is due to electronic transitions from the valence to the conduction bands.
The inset of Fig.\,\ref{fig.optics}\,(c) highlights the frequency range, where these interband transitions set in, and this onset defines the size of the optical gap.

The shape of the loss function at high energies is very similar for all the studied samples, as the values are strongly increasing above $\sim$5000\,\cm\ [Fig.\ \ref{fig.optics} (b)].
As expected, the loss function for the various compounds differs the most in the low-frequency range, where we find a plasmon peak located at different frequency positions. From the position of the plasmon peak, we can determine the value of the screened plasma frequency $\omega_{\mathrm{pl}}^{\mathrm{scr}}$ by fitting the plasmon peak with a Lorentz function, as illustrated in the inset of Fig.\ \ref{fig.optics} (d). The so-obtained values of $\omega_{\mathrm{pl}}^{\mathrm{scr}}$ are plotted in Fig.\,\ref{fig.optics}\,(d) as a function of Sb content $x$.
The pure compound MBT shows the highest value $\omega_{\mathrm{pl}}^{\mathrm{scr}} \approx$1150\,\cm. For the $x$=0.26 compounds, the values are much smaller and there is also a certain difference between the three samples with this composition, as the values are varying between 239 \,\cm (SL3B1) and 459\,\cm (SL3E). The 39\,\% doped sample has the second lowest value $\omega_{\mathrm{pl}}^{\mathrm{scr}}$$\approx$285\,\cm, which indicates, that the $x$=0.39 and the SL3B1 sample have the weakest metallic character. For Sb doping levels above $x$=0.39, the values are increasing steadily with increasing Sb doping ratio up to 768\,\cm\ for the $x$=0.93 sample. These values are in good agreement with predictions from Chen et al.~\cite{Chen.2019}, and are in agreement with the reflectance data, since the weakest metallic properties are found for the 26\,\% and, interestingly, also for the 39\,\% Sb doped samples, while the free charge carrier contributions are higher for lower ($x$$<$0.26) and higher ($x$$>$0.39) Sb doping levels.

It is important to note that even though an insulating state is predicted for the doping level x=0.26 by theoretical calculations~\cite{Chen.2019}, our results are in good agreement with the experimental findings by Lee {\it et al.}~\cite{Lee.2021}, where transport and Hall resistivity measurements of several 26\,\% doped samples have confirmed the presence of either negative or positive free charge carriers.
Hints for a metallic behaviour are found in the optical data for all of our samples, including the three 26\,\% and the 39\,\% Sb doped samples. Since the SL6C and the SL3E sample show a similar optical spectrum and, on the other hand, the SL3B1 and the 39\,\% doped sample resemble each other, we assume that the chemical potential is in first case closer to the conduction bands and in the latter closer to the valence bands, as described by Lee {\it et al.}~\cite{Lee.2021}. Reportedly, samples with 26\,\% Sb doping can be either slightly electron or hole doped, even if they are from the same batch, and ARPES measurements on various samples show that small doping differences at the surface of weak metallic bulks could cause large changes in the Fermi level at the surface~\cite{Lee.2021}. Although IR spectroscopy is mainly bulk and not surface sensitive, a certain contribution from surface states to the free charge carrier response has been reported by Xu~\cite{Xu.2021} and Regmi {\it et al.}~\cite{Regmi.2020}.

\begin{figure*}[t]
	\includegraphics[width=0.9\linewidth]{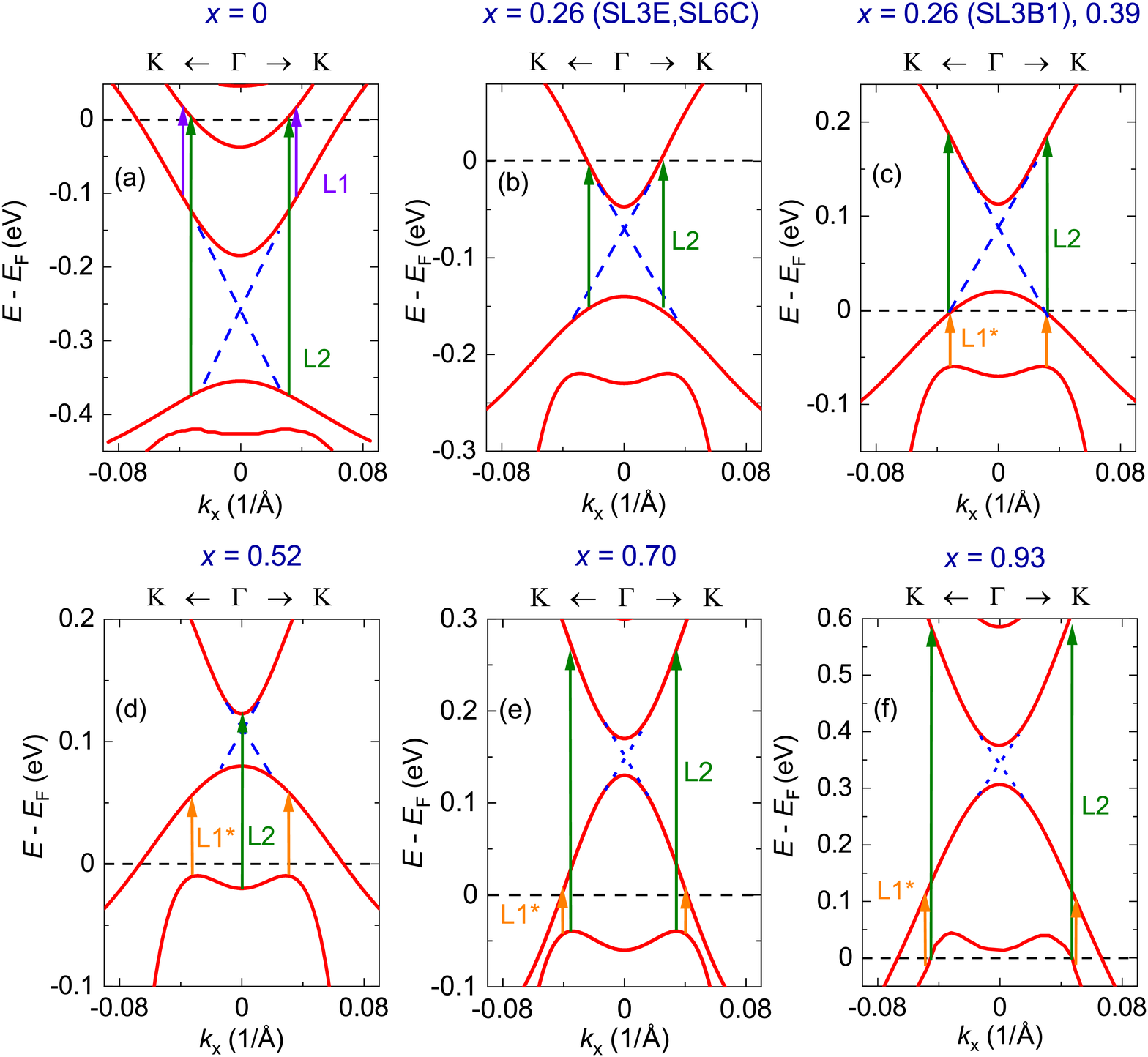}
	\caption{\label{fig.BandStructure} Electronic band structure scheme of MBST for the dopings (a) $x=0$, (b) $x=0.26$ (SL3E,SL6C), (c) $x=0.26$ (SL3B1), $x=0.39$, (d) $x=0.52$, (e) $x=0.70$ and (f) $x=0.93$. Electronic transitions are indicated by violet (L1), orange (L1*), and green (L2) arrows. The topological surface states are displayed with blue dashed and dotted lines. These sketches have been created in accordance with Refs.\ \cite{Chen.2019a,Chen.2019}.}
\end{figure*}

For further analysis, we carried out a simultaneous fit of the reflectivity and optical conductivity spectra
with the Drude-Lorentz model. For the compositions $x=0$ and $x=0.93$, we have implemented two Drude terms, as for these MBST compounds free charge carrier excitations from two bands are expected and two Drude terms (a broad and a narrow one) are also needed to achieve a reasonable fit, as previously shown by us~\cite{Koepf.2020, Koepf.2022}. The scattering rates $\gamma$=1/$\tau$ of the two Drude terms amount to 753~cm$^{-1}$ and 69~cm$^{-1}$ for $x$=0, and 794~cm$^{-1}$ and 96~cm$^{-1}$ for $x$=0.93.
In the case of the compounds with Sb doping x=0.26, 0.39, 0.52 and 0.70, the inclusion of only one Drude term was sufficient to obtain a good fit quality and the fit quality did not improve by implementing two Drude contributions. Hence, for these doping levels, only one electronic band is expected to cross the Fermi level, which was taken into account in the electronic band structure schemes discussed later.
Also for the ``critical'' doping $x=0.26$, where the transition from $n$-type to $p$-type charge carriers takes place, a Drude term was implemented, as Hall resistivity measurements confirmed the presence of free charge carriers for several 26\,\% samples in the studies by Lee {\it et al.}~\cite{Lee.2021}.
Regarding the interband excitations, we have used six or seven Lorentz terms in the measured range, depending on their Sb doping level.
For the samples labelled SL3E and SL6C with $x=0.26$, the L1 oscillator is absent, which will be explained later by the electronic band structure scheme. For sample SL3B1 with $x=0.26$ and for dopings $x\geq 0.39$, L1* replaces the L1 oscillator to indicate a different physical meaning (see below).
The high-frequency contribution labelled ``High Energy'' represents the sum of all higher-energy Lorentz oscillators which lie out of the measured range.
In Figs.\,\ref{fig.fitting}\,(a), (b), and (c) the optical conductivity, the corresponding fit, and all the contributions are shown exemplarily for the samples $x=0$, $x=0.26$ (SL3E) and $x=0.93$, respectively.
One notices that the Drude contributions are affected strongly by the doping level $x$, as expected from the reflectance data.
This also holds for the frequency positions of the low-frequency Lorentz oscillators L1, L1* and L2, as we assume that the substitution of Bi by Sb mostly affects the electronic bands close to the Fermi level.

In Figs.\,\ref{fig.fitting}\,(d) and (e), the plasma frequency $\omega_{\mathrm{pl}}$ and  the zero-frequency conductivity $\sigma_{\mathrm{dc}}$ are compared for all studied samples. In the case of samples $x=0$ and $x=0.93$ with two Drude terms, the effective values of $\omega_{\mathrm{pl}}$ and $\sigma_{\mathrm{dc}}$
have been calculated according to the equations \cite{Fox.2001}
\begin{equation}
\omega_{\mathrm{pl}}=\sqrt{\omega_{\mathrm{pl,1}}^2+\omega_{\mathrm{pl,2}}^2}
\end{equation}
\begin{equation}
\sigma_{\mathrm{dc}}=\frac{\omega_{\mathrm{pl,1}}^2}{4\pi\gamma_1} + \frac{\omega_{\mathrm{pl,2}}^2}{4\pi\gamma_2} \quad ,
\end{equation}
where $\gamma_{1,2}$ represent the scattering rates of the Drude terms obtained by the Drude-Lorentz fits. Furthermore, we depict the positions of the Lorentz oscillators L1, L1* and L2 in Fig.\,\ref{fig.fitting}\,(f) as a function of doping level $x$.
These three oscillators can be ascribed to the low-energy excitations close to the Fermi energy, as described in more detail below.

The (unscreened) plasma frequency $\omega_{\mathrm{pl}}$, which reflects the charge carrier density, shows a similar Sb doping dependence like the screened plasma frequency $\omega_{\mathrm{pl}}^{\mathrm{scr}}$ [see Fig.\,\ref{fig.optics}\,(d)].
The pure compound has the largest $\omega_{\mathrm{pl}}$ value according to its strongest metallic character, while the 26\,\% and 39\,\% doped samples show the smallest values.
In contrast to $\omega_{\mathrm{pl}}^{\mathrm{scr}}$, the values for $\omega_{\mathrm{pl}}$ are not increasing monotonically with increasing $x$, since the 70\,\% Sb doped sample has a smaller plasma frequency than the 52\,\% Sb doped sample, as also revealed by the lower reflectance value at low frequencies [see Fig.\,\ref{fig.optics}\,(a)]. Thus, we find a deviation here in accordance with the predictions of Chen et al.\,\cite{Chen.2019}.
The $\sigma_{\mathrm{dc}}$ values, which also reflect the metallic character, show a similar doping dependence, since one observes a decrease to $x=0.26$, followed by a rather constant behaviour up to $x=0.70$ and an increase to $x=0.93$.
Yet, all doped samples show a weaker metallic character compared to the pure compound according to the values of $\omega_{\mathrm{pl}}$ and $\sigma_{\mathrm{dc}}$.
Regarding the frequency position of the Lorentz oscillators L1/L1* and L2 as a function of doping, we find a similar dependence on the doping level $x$: The undoped sample shows the highest frequencies for both oscillators, followed by a minimum for the ``critical'' doping and a monotonic increase with increasing $x$. The L1 oscillator is attributed to electronic transitions in the pure compound, whereas the L1* oscillator describes transitions in the doped samples, as discussed in detail in the following.

For an interpretation of the low-energy excitations in the Mn(Bi$_{1-x}$Sb$_x$)$_2$Te$_4$ compounds, in particular the low-energy oscillators L1/L1* and L2, we sketch in  Fig.\,\ref{fig.BandStructure} the electronic band structure close to the energy gap and the Fermi energy $E_F$ for the (a) pure, (b) $x$=0.26 (SL3E,SL6C) (c) $x$=0.26(SL3B1)/0.39, (d) $x$=0.52, (e) $x$=0.70, and (f) $x$=0.93 compound, based on Refs.\ \cite{Chen.2019a,Chen.2019}.
According to the different low-energy reflectance spectra, we distinguish between the SL3E/SL6C (x=0.26) and the SL3B1($x$=0.26)/$x$=0.39 samples [see Fig.\,\ref{fig.optics} (a)]. In particular, we assume that the first ones are slightly $n$-doped, whereas the latter ones are slightly $p$-doped, in agreement with the findings by Lee et al.\,\cite{Lee.2021}.
According to the band structure scheme, two Drude terms related to the two bands crossing $E_F$ are expected for the samples $x=0$ and $x=0.93$, whereas one Drude contribution should be present for the other studied samples (i.e., one band crossing $E_F$), in agreement with our optical data analysis.
For all the six highlighted cases in Fig.~\ref{fig.BandStructure}, the low-energy electronic transitions are indicated by vertical arrows, which lead to the Lorentz contributions L1/L1* and L2 in the optical conductivity spectra (see Fig.\,\ref{fig.fitting}). Both the L1 and L1* oscillator describe transitions between two conduction bands [violet (L1) and orange (L1*) arrows], whereas the L2 term (green arrows) is related to transitions across the optical gap, causing a strong onset of spectral weight in the optical conductivity spectrum. In the case of the SL3E and SL6C samples [see Fig.\,\ref{fig.BandStructure}\,(b)], we have not implemented the L1/L1* oscillator, as there is no expected transition between two subbands in contrast to the other dopings, and this is in agreement with the Drude-Lorentz fit [see Fig.\,\ref{fig.fitting}\,(b)].

According to Ref.\ \cite{Chen.2019}, with increasing Sb doping not only $E_F$ is changing its position, but also the energy gap is decreasing, with a reported minimum for $x=0.55$. Therefore, also the optical gap, which is larger than the energy band gap for all studied compounds, is affected by this evolution. As for $x$=0 the Fermi level is energetically quite distant from the energy band gap and at the same time at its maximum for the whole doping series, the optical gap is maximum for this composition. With the closing of the energy gap and the shift of $E_F$ to lower energies with increasing $x$, the optical gap is also decreasing, as the energetic difference between the occupied and unoccupied states of different bands is reducing. For even higher doping levels $x$, the Fermi level is shifting further into the former valence bands, whereby the energy difference to the former conduction band is increasing, which also affects the optical gap in the same way. This behaviour is demonstrated by the frequency shift of L2 with increasing doping. As mentioned above, L1 and L1* describe the transition between two conduction bands; however, since the Fermi level has been shifted to the former valence bands for $x>0.26$, L1 changes its character, indicated by the changed nomenclature (L1*). The energy difference of L1 and L1* of approx.\ 80\,meV is thus justified, since the energetic difference of the involved electronic bands in the L1 transition is larger compared to those of L1* (see Fig.\,\ref{fig.BandStructure}). In comparison, the frequency shift of the L1* oscillator with doping is quite weak, still a steady increase to higher values with increasing $x$ can be detected.

\section{Conclusion}
In conclusion, we studied the optical response of Mn(Bi$_{1-x}$Sb$_x$)$_2$Te$_4$ samples with various Sb doping levels $x$ by reflectance measurements over a broad frequency range, in order to observe the effect of the gradual exchange of Bi by Sb on the low-energy electronic excitations. In good agreement with recent studies~\cite{Chen.2019,Yan.2019a,Lee.2021}, the metallic character is affected by the Sb doping: it is weakened from $x=0$ to $x=0.26$, where an insulating composition has been predicted. For doping levels above $x=0.39$ the metallic character increases for increasing Sb content up to the highest measured doping level $x=0.93$.
As indicated by the presence of at least one Drude term, all studied samples show a metallic character.
The high-energy part of the optical response is only weakly affected by the doping level, which signals that mainly the electronic band structure in close vicinity to the Fermi energy is sensitive to Sb doping. Different $x=0.26$ crystals show significant differences in the profile of their reflectance spectra. Furthermore, a strong position dependence of the optical response is found for $x=0.26$ in contrast to other doping levels. We propose a scheme of the electronic band structure close to $E_F$ as a function of the doping level, which explains our optical results.

\begin{acknowledgments}
C.\ A.\ K.\ acknowledges financial support from the Deutsche
Forschungsgemeinschaft (DFG), Germany, through Grant
No. KU 1432/15-1.  Z.Q.M. and S.H.L. acknowledges the support of the US NSF through the Penn State 2D Crystal Consortium-Materials Innovation Platform (2DCC-MIP) under NSF Cooperative Agreement DMR-2039351.
\end{acknowledgments}

\end{document}